\documentclass{epl}

\usepackage{graphicx}

\def\mys#1{{\mbox{\scriptsize{#1}}}}    % font for capital subscripts

\def\gp{g_{\mys{P}}}             % number of monomers in a pearl
\def\KB{k_{\mys{B}}}            % Boltzmann factor
\def\lb{\ell_{\mys{B}}}         % Bjerrum length
\def\np{n_{\mys{P}}}             % number of pearls
\def\NM{N_{\mys{m}}}            % Chain length
\def\rpp{r_{\mys{PP}}}          % pearl-pearl distance
\def\RE{R_{\mys{E}}}            % End-to-end distance
\def\RG{R_{\mys{G}}}            % Radius of gyration
\def\rhom{\rho_{\mys{m}}}       % Monomer density
\title{Structure of Polyelectrolytes in Poor Solvent}
\shorttitle{Polyelectrolytes}

\author{Hans J{\"o}rg Limbach\inst{1} \and Christian Holm\inst{1} \and Kurt Kremer\inst{1}}

\institute{
  \inst{1} Max-Planck-Institut f\"{u}r Polymerforschung, PO Box 3148, 55021 Mainz, Germany
}


\begin{document}

\maketitle

% old abstract
%\begin{abstract}
%  We present simulations on charged polymers in poor solvent.  First we
%  investigate in detail the dilute concentration range with and without
%  imposed extension constraints.  The resulting necklace polymer conformations
%  are analyzed in detail.  We find strong fluctuations in the number of pearls
%  and their sizes leading only to small signatures in the form factor and the
%  force-extension relation. The scaling of the peak in the structure factor
%  with the monomer density shows a pertinent different behavior from good
%  solvent chains.
%\end{abstract}

\begin{abstract}
  We present molecular dynamics simulations on strongly charged
  polyelectrolytes in poor solvent. The resulting pearl-necklace
  conformations are analyzed in detail. Fluctuations in the number of
  pearls and their sizes lead only to small signatures in the form factor
  and the force-extension relation, which is a severe obstacle for
  experimental observations. We find that the position of the first
  peak in the structure factor varies with the monomer density as
  $\approx \rhom^{0.35}$ for all densities. This is a qualitative
  difference to polyelectrolyte solutions in good solvent which scale
  as $\rhom^{1/3}$ and $\rhom^{1/2}$ in the dilute and semi-dilute
  concentration regime, respectively.
\end{abstract}

%====================================================================
%   Introduction
%====================================================================

Polyelectrolytes (PEs) are polymers that carry ionizable groups that
dissociate ions in aqueous solution.  Technical applications range
from viscosity modifiers, precipitation agents, superabsorbers to leak
protectors~\cite{hara93a}.  In biochemistry and molecular biology they
are of great importance because virtually all proteins, as well as
DNA, are PEs.

Many PEs contain a non-polar hydrocarbon backbone, for which water is
a poor solvent.  Therefore, in aqueous solution, there is a
competition between the tendency to precipitate, the Coulomb
interaction and the entropic degrees of freedom.  This can lead to
elongated strings of locally collapsed structures (pearls).  Such
necklace conformations have been predicted on the basis of scaling
arguments in ref.~\cite{scaling} for a weakly charged single chain PE,
and have been confirmed by simulations using the Debye-H\"uckel
approximation~\cite{mcsims,scaling} and with explicit
counterions~\cite{micka99a}.  The formation of the necklace
structure is due to the Rayleigh instability of a charged droplet,
which leads to a split once a critical charge is reached.  The size of
the pearls is determined by the balance between electrostatic
repulsion and surface tension. The distance between two pearls is
governed by the balance of the electrostatic pearl-pearl repulsion and
the surface tension.
However, there is up to now no clear experimental proof for the
existence of necklace chains~\cite{carbajal00a,heinrich01a}.
Our previous papers\cite{micka99a} dealt with smaller system
sizes and showed that necklaces exist also in the presence of
counterions and exhibit a variety of conformational transitions as a
function of density.  The focus of the present paper is to analyze by
extensive computer simulations in detail three possible experimental
observables, namely the form factor, the structure factor and the
force-extension relation, which can be probed by scattering and AFM
techniques.
Analyzing the fluctuations of necklace structures we find an extended
coexistence regime between different necklace structures and broad
distributions for the pearl sizes and the pearl-pearl distances that
smear out the necklace signatures. In addition we find that the peak
in the structure factor of the solution scales proportional to the
density $\rhom^{1/3}$ from the dilute up into the dense phase which
is in clear contrast to good solvent PEs.
%

%====================================================================
% Simulation Method
%====================================================================

Our PE model is the same as in Ref. \cite{micka99a} and consists of a
bead spring chain of Lennard-Jones particles, whose interaction at
distance $r < R_c$ is given by
$4\,\epsilon[(\frac{\sigma}{r})^{12}-(\frac{\sigma}{r})^{6}-c]$, and
is zero elsewhere.  The constant $c$ is chosen such that the potential
value is zero at the cutoff $R_c$, and $\epsilon$ is a measure of the
solvent quality.
Chain monomers interact up to a distance $R_c = 2.5 \sigma$ and
experience an attraction with $\epsilon=1.75 \, \KB T$. The
$\Theta$-point for this model is at $\epsilon=0.34 \KB
T$~\cite{micka99a}.
The counterions interact via a purely repulsive interaction with
$R_c=2^{1/6}\sigma$.  The units of length, energy and time are
$\sigma$, $\epsilon$, and $\tau$, respectively.
For bonded monomers we add a bond potential of the form $- 14 \KB T
\ln(1- (\frac{r}{2\sigma})^2)$, which results in an average bond
length $b = 1.09 \sigma$.
Charged particles at separation $r_{ij}=|{\bf r_j - r_i}|$ interact
via the Coulomb energy $\KB T \lb q_i q_j / r_{ij}$, with $q_i =
1,(-1)$ for the charged chain monomers (counterions). The Bjerrum
length $\lb=e^{2}/(4\pi\epsilon_{S}\epsilon_{0}\KB T)$ ($e$:
unit charge, $\epsilon_{0}$ and $\epsilon_S $: permittivity of the
vacuum and of the solvent) is set to $\lb=1.5\sigma$.
A velocity Verlet algorithm with a standard Langevin thermostat is
used to integrate the equation of motion~\cite{grest86a} (friction
coefficient $\Gamma = \tau^{-1}$, time step $\Delta t=0.0125\tau$).
%After equilibrating we measured for about $10^7$ time steps in each
%case. 
The Coulomb interaction was calculated with the
P3M-algorithm~\cite{deserno98}, tuned to force accuracies well above
the thermal noise level.  The central simulation box of length $L$
contained $N_P=5$, $10$ or $32$ chains subject to periodic boundary
conditions.  Each chain consisted of $\NM= 48 \dots 478$ monomers,
with a charge fraction $f=1/3 \dots 1/2$.  Other parameter sets yield
qualitatively similar results~\cite{limbach01c,limbach02b}, so that we
restrict ourselves here to the above described ``generic'' system.
The simulation time for the longest chains was around $200$ times the
measured correlation time for the end-to-end distance $\RE$ and the
chains centers of mass diffused at least several radii of gyration
$\RG$.  The pressure $p$ is always positive and the $p V$ diagram
is convex at all densities, thus our simulations are
stable, reach true thermal equilibrium, and reside in a one phase
region.
%Especially they
%are not trapped in meta-stable states, which in some cases we also
%checked by using different initial conditions.  
The volume density inside the pearls does not exceed $0.47$, which is
well below the glass transition.

%====================================================================
% Results
%====================================================================

\begin{figure}
\onefigure[width=0.65\linewidth]{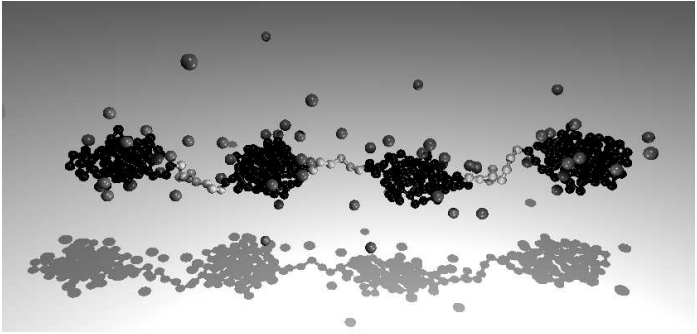}
\caption{Snapshot of a necklace structure analyzed with the cluster
  algorithm. Chain length $\NM=382$: (black) monomers in pearls,
  (light gray) monomers in strings, (gray) counterions.}
\label{fig.pnsnap}
\end{figure}

\begin{figure}
\onefigure[width=0.55\linewidth]{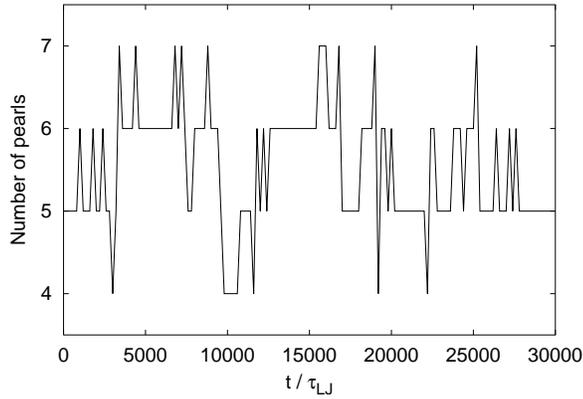}
\caption{Time evolution of a single chain of length
  $\NM=430$ in equilibrium, shown over a quarter of the physical
  running time. Many transitions between different structure
  types can be observed.}
\label{fig.trace}
\end{figure}

Being first interested in single chain properties we used a monomer
density $\rhom=\frac{N_P \NM}{L^3}=1.5\times 10^{-5}\sigma^{-3}$. A typically 4-pearl
(and 3-string) structure is shown in fig.~\ref{fig.pnsnap}. The
substructures contain 90-8-94-6-77-9-98 monomers.  For the
identification of the pearls and strings we used an especially adapted
cluster algorithm\cite{limbach01c,limbach02b}.
%
% figure 1
% Snapshot of an analyzed  necklace structure 
%
% Coexistence of different structure types / Fluctuations
%====================================================================
%
For chains of length $\NM=430$, we find structures with 4, 5, 6 and 7
pearls and an average pearl number $\np=5.22$. Results for other chain
lengths can be found in table~\ref{tab.basdat}.
%
% Basic observables for different chain length
\begin{table}
\caption{Basic observables for various chain lengths at
$\rhom=1.5\times 10^{-5}\sigma^{-3}$. 
Symbols: Chain length $\NM$, end-to-end distance $\RE$,
average number of pearls $\np$, distance between neighboring pearls
$\rpp$, number of monomers $\gp$ in a pearl, pressure $p$ and the
osmotic coefficient $\Pi$. Types denotes the observed structures with
$\np$ pearls, and $\delta \rpp$ and $\delta \gp$ denote the relative
standard deviations. All systems had a Bjerrum-length $\lb=1.5
\sigma$ and a charge fraction $f=1/3$. 
  \label{tab.basdat}}
\begin{center}
\begin{tabular}{cccccccccccc}
$\NM$&$\RE$ [$\sigma$]& $\np$&Types &$\rpp[\sigma]$&$\delta \rpp$&$\gp$&$\delta \gp$&$p$[$10^{-6} \KB T \sigma^{-3}$]&$\Pi$ \\ \hline\hline
46   & 3.48          & 1.0  & 1       & -    & -    & 48   & 0.0 &4.79&0.96\\%PS0030
142  & 12.9          & 1.97 & 2       & 11.6 & 0.16 & 67.5 & 0.17&2.97&0.59\\%PS0032
238  & 25.4          & 2.89 & 2 - 4   & 12.8 & 0.20 & 75.5 & 0.30&2.31&0.46\\%PS0033
334  & 30.4          & 3.94 & 3 - 5   & 13.1 & 0.21 & 78.6 & 0.33&2.0 &0.40\\%PS0035
382  & 45.4          & 4.53 & 3 - 6 & 13.3 & 0.22 & 78.0 & 0.32  &1.75&0.35\\%PS0004
430  & 55.4          & 5.22 & 4 - 7 & 13.2 & 0.22 & 75.3 & 0.35  &1.78&0.36\\%PS0036
478  & 79.7          & 5.60 & 4 - 7 & 13.3 & 0.21 & 78.3 & 0.34  &1.76&0.35\\%PS0037
\end{tabular}
\end{center}
\end{table}
The coexistence regime between different structure types enlarges with
increasing chain length. One reason for the small differences in the
free energy is the interplay between the chain conformation and the
counterion distribution.  Conformations with a lower number of pearls
have a smaller extension, e.g.  $\RE$, and a larger pearl size. This
leads to a stronger attraction of counterions and yields a smaller
effective charge on the chain which in turn stabilizes larger pearl
sizes and smaller chain extensions.  In contrast to scaling
theories~\cite{khokhlov80a,schiessel98a,dobrynin99a,vilgis00a} we do
not find a collapse into a globular state due to this effect. This is
consistent with a more refined theoretical analysis \cite{deserno01f}
that takes prefactors and finite concentrations into account. The
interplay between counterion distribution and structural changes of
the conformation can be measured e.g. via the number of counterions
$n_c(n)$ for a $n$ pearl structure inside a shell of $3\sigma$ from
the chain.  For example, for $\NM = 430$ we find $n_c(4)=60.4$, $n_c(5)=56.7$,
$n_c(6)=52.8$ and $n_c(7)=50.2$.
% from PS0036/iondis.dat
% Ions inside 3 sigma: P = 55.9, P(4) = 60.4 P(5) = 56.7 P(6) = 52.8
% P(7) = 50.2 
From the probability $p(n)$ of finding a $n$ pearl structure we 
calculate the free energy difference $\Delta {\cal F}^{nm} = \KB T \ln
\left( p(n)/p(m) \right)$.  For $\NM=430$ we find $\Delta {\cal
  F}^{45} = -1.33\KB T$, $\Delta {\cal F}^{56} = 0.66 \KB T$, $\Delta
{\cal F}^{67} = 1.90\KB T$. All values are of the order $\KB T$ which
is consistent with the observed large coexistence regime.
% dataset from PS0036
% p4=0.1408 p5=0.5348 p6=0.2766 p7=0.0414
In order to exclude the possibility of metastable states, we have
observed the time evolution for a single chain. We find typically many
transitions between different structure types, see i.e. one example
for $\NM=430$ in fig.~\ref{fig.trace}.
%
% figure 2
% Time development of the structure type for a chain 
Not only the pearl number, also the position and size of the pearls
fluctuate strongly. For chains with more than 2 pearls, both the mean
values as well as the relative standard deviations $\delta$ of the
distance $\rpp$ between neighboring pearls and the number $\gp$ of
monomers in a pearl are almost constant, see table~\ref{tab.basdat},
which is consistent with scaling theory.

%   Form-factor
%====================================================================

Scattering experiments probe the conformation via the chain form
factor $S_1(q)$ given by
\begin{equation}
S_1(q) = \frac{1}{4 \pi \NM} \left< \sum_{i,j=1}^{\NM} 
         \frac{\sin(q \, r_{ij})}{q \,r_{ij}} \right>,
\label{eq.formfac} 
\end{equation}
where $\left<\dots\right>$ denotes the ensemble average.
%
% figure 3
% Form factor
\begin{figure}
\onefigure[width=0.65\linewidth]{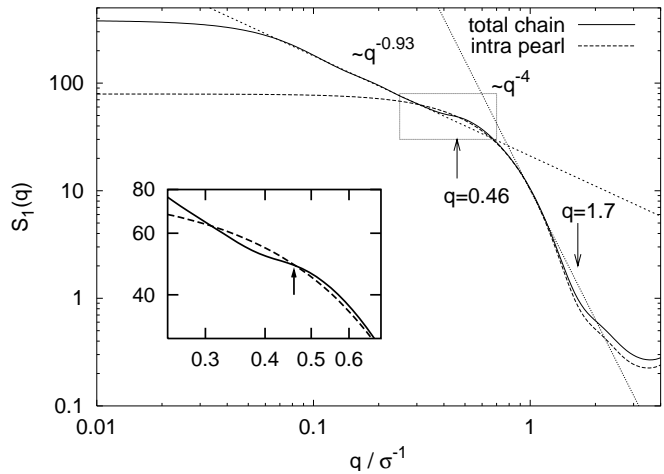}
\caption{Form-factor $S_1(q)$ for pearl necklaces: (solid) form-factor
  of the whole chain, (dashed) form-factor due to intra pearl
  scattering, (dotted) fits. The marked region is enlarged in the inset.}
\label{fig.formfac}
\end{figure}
%
% Some math for the form factor:
% 1) Inter Pearl scattering: The pearl pearl distance can be
%    calculated from the inflection point: r_PP = (2 pi)/q_i
% 2) Intra Pearl scattering: connection between minimum and pearl
%    radius: (2 pi)/(q_m*1.4); q=1.7 gives a Pearl radius of 2.6
Fig.~\ref{fig.formfac} shows the measured form factor for
$\NM=382$ (see table~\ref{tab.basdat}).  In the Guinnier regime ($\RG
q \ll 1)$ the radius of gyration $\RG$ can be calculated from
$S_1(q)=\NM \left(1 - (\RG q)^{2}/3\right)$, giving $\RG=16.8\sigma
\pm0.3\sigma$, in agreement with the directly calculated value
$\RG=16.9\sigma\pm0.4\sigma$. In the range $0.07\sigma^{-1} \le q \le
0.3\sigma^{-1}$ $S_1$ scales as $q^{-0.93}$ which corresponds to a
stretched object. Around $q=0.46\sigma^{-1}$ the form factor has an
inflection point. A comparison with the intra pearl scattering reveals
that this is due to inter pearl scattering (inset in
fig.~\ref{fig.formfac}).  Dividing out the intra pearl form factor
gives access to the inter pearl scattering and thus $\rpp$. The form
factor yields $\rpp=13.6\sigma$, in accord with the directly measured
value $\rpp=13.3\sigma$ (table~\ref{tab.basdat}).  In the region
around $q=1.0 \sigma^{-1}$ we find $S_1(q) \propto q^{-4}$, the
typical Porod scattering. From the small dip at $q=1.7 \sigma^{-1}$
one can calculate the radius $r_P$ of the pearls to be $r_P=2.6\sigma$
which again compares well to the real space value $r_P\simeq 3\sigma$.
We conclude that the cooperative effect of fluctuations on overlapping
length scales broadens all characteristic signatures which can be
revealed by scattering under experimental conditions (polydispersity,
charge fluctuations, etc). Thus necklaces might be difficult to
detect.

% Density dependence of the correlation length
%====================================================================

\begin{figure}
\onefigure[width=0.65\linewidth]{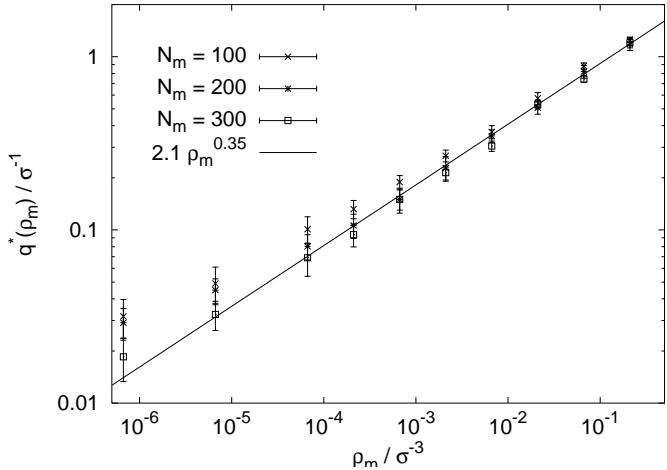}
\caption{Density dependence of the peak $q^*$ in the structure factor
  for three different chain length $\NM=100, 200, 300$ with $f=0.5$.
  The black line is a fit to the data with $\NM = 200$.}
\label{fig.correl}
\end{figure}

The overall scattering function $S(q)$ of the solution contains
additional experimental information. We analyze here the inter chain
scattering $S_{IC} = S/S_1$. For good solvent PEs
experiments~\cite{nierlich79a}, theory~\cite{joanny01a}, and
simulations~\cite{stevens95a} find a pronounced first peak of $S_{IC}$
at $q^* = (2\pi) / \xi$, where $\xi$ is the correlation length. The
position varies as $q^* \propto \rhom^{1/3}$ in the very dilute
regime and crosses over to a $\rhom^{1/2}$ regime at higher
concentrations.  In fig.~\ref{fig.correl} we have plotted the density
dependence of $q^*$ in poor solvent for different chain lengths.
%
% figure 4
% Density dependence of the correlation length
Within the error bars we find that for poor solvent chains $q^*$
scales proportional to $\rhom^{0.35\pm0.04}$ for {\it all}
concentrations and chain lengths.
Possible reasons for this are that the response
of the polyelectrolyte conformation to density changes is much larger
in the poor solvent case~\cite{micka99a,limbach01c} than in the good
solvent case~\cite{stevens95a}, and the chain extension behaves
non-monotonic as a function of density~\cite{micka99a,limbach01c}.
Furthermore, in the density regime between $\rhom=10^{-2}\sigma^{-3} \dots
10^{-4}\sigma^{-3}$ the chain
extension and the pearl number varies most, and
almost all monomers are located within the pearls. 
Upon approaching
the dense regime, the string length tends to zero and we find a chain
of touching pearls, indicating that the conventional necklace picture
breaks down. Our result is compatible to scaling exponents found in
scattering experiments~\cite{peakscaling}.

Scaling theories\cite{dobrynin99a,dobrynin01a} have predicted a
$\rhom^{1/2}$ regime to start at $\rho^*_o$, which is defined at the
density where $\RE \approx \xi$, and to extend until $\xi \approx
\rpp$ where a bead-controlled $\rhom^{1/3}$ regime starts. We find
$\rho^*_o \simeq 5\times10^{-2}, 10^{-3}, 10^{-4}$ for $\NM = 100,
200, 300$, and $\rpp$ is equal to $\xi$ between $\rhom=10^{-2}$ and
$10^{-1}$.  One probable reason for our different findings is the
strong inter-chain coupling and influence of the counterions on the
conformations which are not sufficiently taken into account in the
mean field approach.  It is not clear at this stage if the
$\rhom^{1/2}$ regime can be recovered for much longer chain length.
In addition we observe that the chains form a transient physical
network at $\rhom=0.2 \sigma^{-3}$ for $\NM \ge 200$ which has
neither been seen in previous simulations nor predicted by theoretical
approaches but is in accord with experimental
studies~\cite{peakscaling}. During the simulation time these networks
reconstruct several times, e.g. chains are not trapped.

\begin{figure}
\onefigure[scale=0.6]{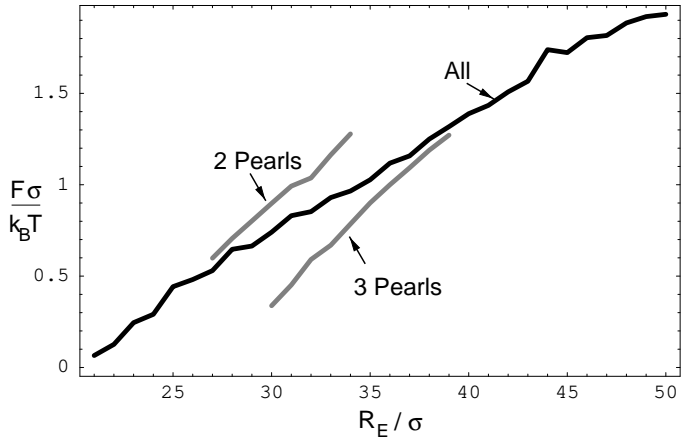}
\caption{Force-extension relation for a transition from two to three
  pearls. Shown are the average over all conformations, over only
  two-pearl configurations and only three-pearl configurations.}
\label{fig.stress}
\end{figure}

% Stress-strain relation
%====================================================================
%
A different way to measure a necklace signature can be obtained by
single-molecule force spectroscopy, e.g. stretching the chain by
AFM~\cite{hugel01a} or optical tweezers. The case of an imposed
end-to-end distance was investigated in
Refs.~\cite{tamashiro00a,vilgis00a,picket01a} for a weakly charged
chain at infinite dilution, for which a saw-tooth pattern in the
force-extension curve was predicted.  We performed simulations in the
weak coupling limit, using a single chain with $\NM=256$, $f=1$,
$\lb=0.08\sigma$, and no counterions present.  At equilibrium this
system is in a two-pearl configuration with $\RE=21\sigma$. The
force-extension curve was obtained by imposing various fixed $\RE$ up
to $\RE=50\sigma$. Then the force on the end-monomers was measured, see
fig.~\ref{fig.stress}.  The first remarkable observation is that the
two-pearl state evolves into a three-pearl state under extension,
which was predicted in ref.~\cite{tamashiro00a,picket01a}. This is
counter intuitive from simple arguments, but can be shown to result
from the electrostatic inter-pearl interactions~\cite{limbach01c}.
%
% figure 5
% Stress strain relation
%%%%%%%%%%%%%%%%%%%%%%%%%%%%
%%%%%%%%%%%%%%%%%%%%%%%%%%%%
%
Only when the force averages where computed separately for the three-
and two-pearl states, we recognize the predicted saw-tooth pattern in
the force.  In equilibrium one would expect a rounded plateau for the
transition, and the upper and lower part would correspond to
metastable superheated and supercooled states.  Again, due to the
large fluctuations all nontrivial signatures in the force-extension
relation are washed out.  Analyzing the variance of the measured
forces by applying a time window of the order of the transition time
we find a $25\%$ increase in the width of the force distribution in
the transition region $\RE=30 \dots 34 \sigma$, and a bimodal force
distribution.  This means that one needs an experimental time
resolution below one $\mu s$ to directly resolve the structure
dependent force difference.  An estimate of typical viscous and
inertial forces acting on colloidal particles shows that the Brownian
motion of small spheres attached to a single PE chain might indeed
reflect the fluctuating forces exerted by the chains. The experimental
determination of theses fluctuating forces by optical
nanorheology~\cite{gisler98a} or AFM noise analysis~\cite{gelbert00a}
presumably is difficult due to the high characteristic frequencies
involved.

%====================================================================
%  Conclusions
%====================================================================

To sum up we find that poor solvent PEs show an extended coexistence
regime between different necklace structures. We observe stable
necklaces even with condensed counterions.  In all investigated cases
the signatures of necklaces in the chain form factor are broadly
smeared out.  A similar broadening of signatures is found for the
force-extension curve which will obscure any of the recently predicted
saw-tooth patterns~\cite{vilgis00a,tamashiro00a,picket01a}.
Nevertheless, the simulations reproduce the predicted initial increase
of pearl number under imposing an external length
constraint~\cite{tamashiro00a,picket01a}.  Our results suggest that
the strong fluctuations impose a severe obstacle in observing necklace
structures and structural transitions directly in experiments.
Finally  the location
of the first peak in the solution structure factor is qualitatively
different from that of PEs in good solvent.  This is in good
accordance to experiments~\cite{peakscaling} but does not show the
full complexity of the proposed scaling picture at this
point~\cite{dobrynin01a}.
\acknowledgments
We thank B. D{\"u}nweg, D. Johannsmann, B. Mergell, H. Schiessel and
M. Tamashiro for many fruitful discussions and comments.  A large
computer time grant hkf06 from NIC J\"ulich is also gratefully
acknowledged.

{\it Note: After the submission of this work we learnt that recent light
  scattering measurements of D. Baigl and C. E. Williams of $q^*(\rhom)$ for PSS
  solutions produce an exponent of 0.38 for a charge fraction of 55\% which is
  in good agreement with our simulations.}

\end{document}